\documentclass{webofc}
\usepackage[varg]{txfonts}   

\usepackage{tikz}
\usetikzlibrary{shapes.arrows}
\usetikzlibrary{positioning}

\begin{document}
\title{Deep learning for flow observables in high energy heavy-ion collisions }
%
%

\author{\firstname{Henry} \lastname{Hirvonen}\inst{1,2}\fnsep\thanks{Speaker, \email{hevivahi@jyu.fi}} \and
        \firstname{Kari} \lastname{J.~Eskola}\inst{1,2}\fnsep\thanks{\email{kari.eskola@jyu.fi}} \and
        \firstname{Harri} \lastname{Niemi}\inst{1,2}\fnsep\thanks{\email{harri.m.niemi@jyu.fi}}
}

\institute{University of Jyv\"askyl\"a, Department of Physics, P.O.B. 35, FI-40014 University of Jyv\"askyl\"a, Finland
\and
           Helsinki Institute of Physics, P.O.B. 64, FI-00014 University of
Helsinki, Finland
          }

\abstract{%
 We demonstrate how deep convolutional neural networks can be trained to predict 2+1 D hydrodynamic simulation results for flow coefficients, mean-$p_T$ and charged particle multiplicity from the initial energy density profile. We show that this method provides results that are accurate enough, so that one can use neural networks to reliably estimate multi-particle flow correlators. Additionally, we train networks that can take any model parameter as an additional input and demonstrate with a few examples that the accuracy remains good. The usage of neural networks can reduce the computation time needed in performing Bayesian analyses with multi-particle flow correlators by many orders of magnitude.  
}
\maketitle
\section{Introduction}
\label{sec:intro}
Neural networks have proven to be an effective tool for a variety of applications in heavy-ion physics. These range from performing pre-processing or selection of large data flows in experiments to emulating computationally expensive simulations \cite{ATLAS:2019bwq, Liu:2022hzd, Wang:2023muv, Hirvonen:2023lqy}. The rising popularity of neural networks is driven by their accuracy and fast inference speed when dealing with complex multi-dimensional data. These aspects can be crucial when performing real-time data selection or heavy numerical simulations that need to be repeated a large number of times. 

The reduced computation time is especially needed when trying to extract the matter properties of the quark-gluon plasma (QGP) from the experimental data through hydrodynamic simulations using a Bayesian analysis. This is due to the fact that one Bayesian analysis will need $\sim 10^6-10^9$ simulated collision events depending on which measured observables one includes in the analysis. Performing this many hydrodynamic simulations will take $\sim 10^5-10^8$ CPU hours, which makes the inclusion of some multi-particle correlations impractical, even though they could provide additional information to constrain the QCD matter properties.  

In principle, all the final state information in hydrodynamic simulation is encoded into the initial state and the matter properties of QGP. However, extracting the final state information directly from the initial state is a highly nontrivial task since relativistic hydrodynamics is a nonlinear theory. The convolutional networks are particularly good at detecting patterns in structured 2-dimensional data, like images, which is why they are excellent tools when trying to estimate the final state observables from an initial state event by event.

\section{Neural network}
\label{sec:implementation}
The convolutional neural networks, one for each observable, are trained to produce $p_T$-integrated flow coefficients $v_n$, mean transverse momentum $[p_T]$ and charged particle multiplicities $dN_{ch}/d\eta$. This was originally done in Ref. \cite{Hirvonen:2023lqy}, from where one can find a detailed description of the setup. Here we just go through the main points. 

As the training data, we used 20 k EKRT (Eskola-Kajantie-Ruuskanen-Tuominen) model ~\cite{Eskola:1999fc, Paatelainen:2012at, Paatelainen:2013eea, Niemi:2015qia} initial energy density profiles in the transverse plane and the corresponding final state observables in these events at midrapidity. The final state observables are obtained from the 2+1 D hydrodynamic simulations done in Ref.~\cite{Hirvonen:2022xfv}. The training events are distributed evenly between four different collision systems: 200 GeV Au+Au, 2.76 TeV Pb+Pb, 5.023 TeV Pb+Pb, and 5.44 TeV Xe+Xe. Even though we use one specific setup of an initial state model combined with a hydrodynamics code to produce the training data, the methods introduced here are expected to be applicable also with training data obtained from any other setup of a similar type. 

The neural network architecture is the DenseNet architecture~\cite{HuangLW16a} with slight modifications that will make it suitable for regression tasks. For a more complete description of the architecture, see Ref.~\cite{Hirvonen:2023lqy}. It took $\sim $ 1 hour to train a network that can produce one observable. With a set of trained networks, one can generate 1 M events in $\sim$ 20 hours.

\begin{figure*}
\centering
\includegraphics[width=\textwidth]{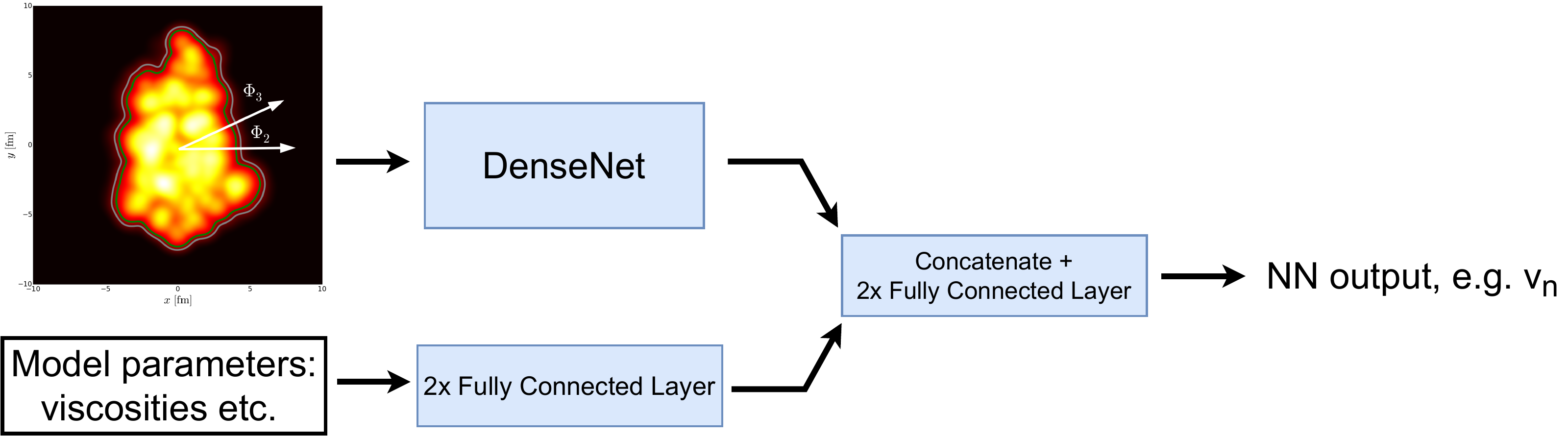}
\caption{Schematic presentation of neural network structure with multiple inputs.}
\label{fig:architecture}       
\end{figure*}

\subsection{Model parameters as an input}
\label{sec:extension}
A neural network that can predict a final state observable from an initial state is already a lot faster than doing full hydrodynamic simulations, but it has a drawback: every time one wants to change QCD matter properties or model parameters that affect time evolution of the system, one would need to generate a new set of training data and retrain the networks. This issue can be solved by adding all the parameters of interest as additional input to the neural networks. Here we refer to this type of network with additional input parameters as $\rm NN_p$. The architecture of $\rm NN_p$ is demonstrated in Fig.~\ref{fig:architecture}. The energy density input is treated the same way as without additional inputs and all the additional inputs are put through two fully connected layers and then combined with the output of the DenseNet layer structure. After this, we have included two fully connected layers from which we then obtain the final output. The training of $\rm NN_p$ was done using in total of 160 k training events distributed evenly between 4 collision systems and 2 k parameter points sampled from a Latin hypercube. This makes only 80 events of training data for one parameter point, which is 250 times more efficient than the training in the previous case.

\begin{figure*}
\centering
\includegraphics[width=\textwidth]{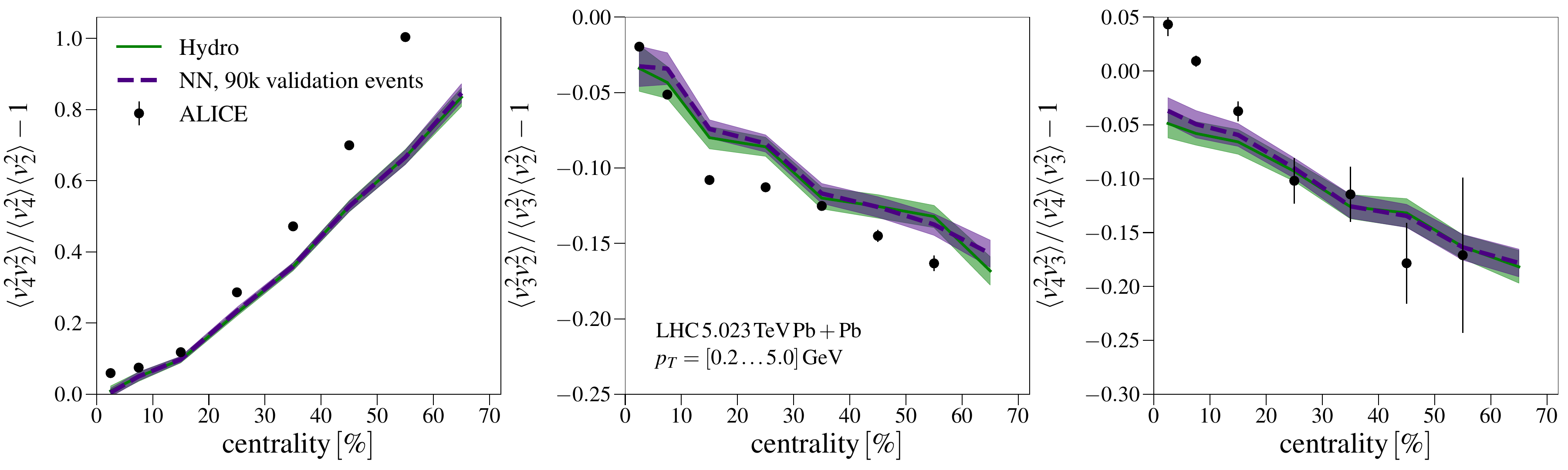}
\caption{Validation test of the neural networks for $NSC(m,n)$ with 90 k validation events. The experimental data are from the ALICE Collaboration \cite{ALICE:2021adw}. Figure from Ref. \cite{Hirvonen:2023lqy}.}
\label{fig:validation}        
\end{figure*}

\begin{figure*}
\centering
\vspace{-5mm}
\includegraphics[width=\textwidth,clip]{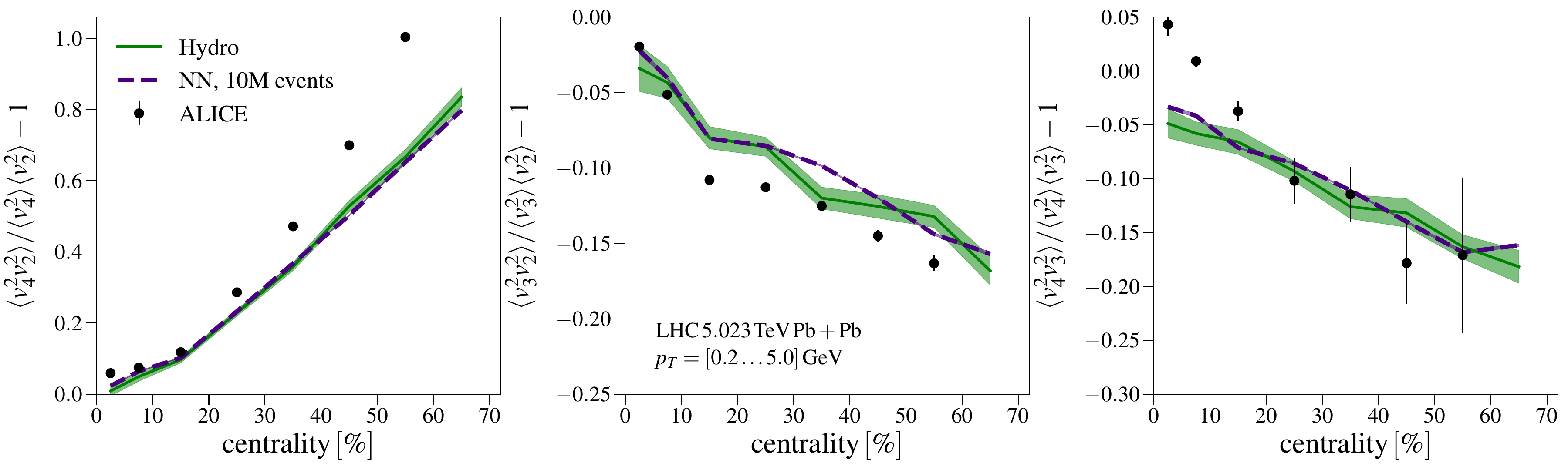}
\caption{Neural network prediction for $NSC(m,n)$ with 10 M events. The experimental data are from the ALICE Collaboration \cite{ALICE:2021adw}. Figure from Ref. \cite{Hirvonen:2023lqy}.}
\label{fig:high_statistics}       
\end{figure*}

\section{Results and conclusions}
\label{sec:results}
To test the accuracy of the neural networks that were trained with one set of model parameters, we generated 90 k independent EKRT initial energy density profiles and compared the results of hydrodynamic simulations against the neural network predictions. In Fig.~\ref{fig:validation} we show a comparison of these two for normalized symmetric cumulants $NSC(m,n)$ as a function of centrality. We can see that the neural network can reproduce the cumulants well, even though the size of the training data for one collision system was only 5 k events. In Fig.~\ref{fig:high_statistics}, we demonstrate how one can then generate 10 M events with the neural networks to see how these cumulants would look when statistical errors became insignificant. Here we can see quite noticeable deviations from the result that used 90 k events, especially for $NSC(2,3)$, for which the centrality dependence clearly changes, matching the shape of the ALICE measurements better. This illustrates the importance of the number of events used when trying to constrain the QCD matter properties with multi-particle flow correlations.  

\begin{figure*}
\centering
\begin{tikzpicture}
    \node (a) at (0.0, 0.0) {\includegraphics[width = 0.25\textwidth]{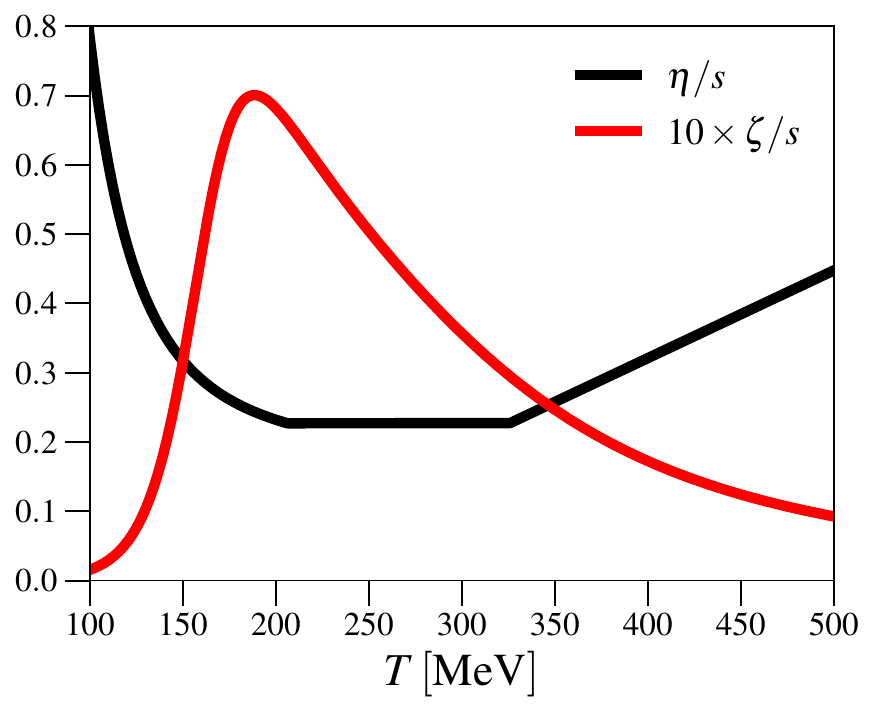}};
    \node (b) at (6.65, 0.0) {\includegraphics[width = 0.715\textwidth]{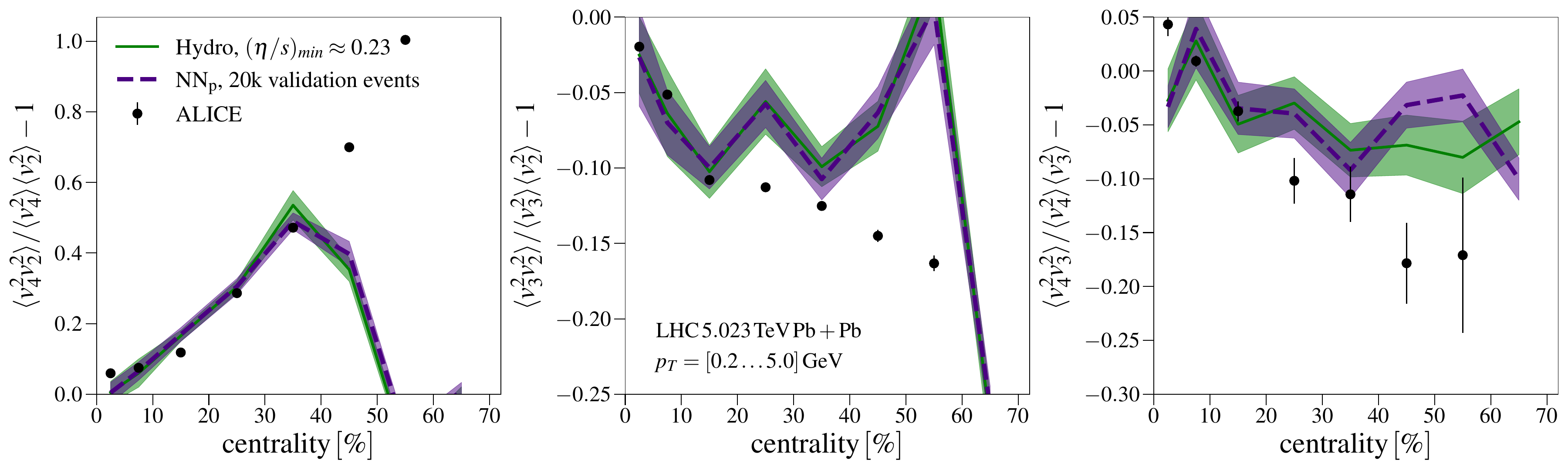}};
    \draw [ultra thick,black,->] (1.55, 0.0) to  (2.0, 0.0);
\end{tikzpicture}

\begin{tikzpicture}
    \node (a) at (0.0, 0.0) {\includegraphics[width = 0.25\textwidth]{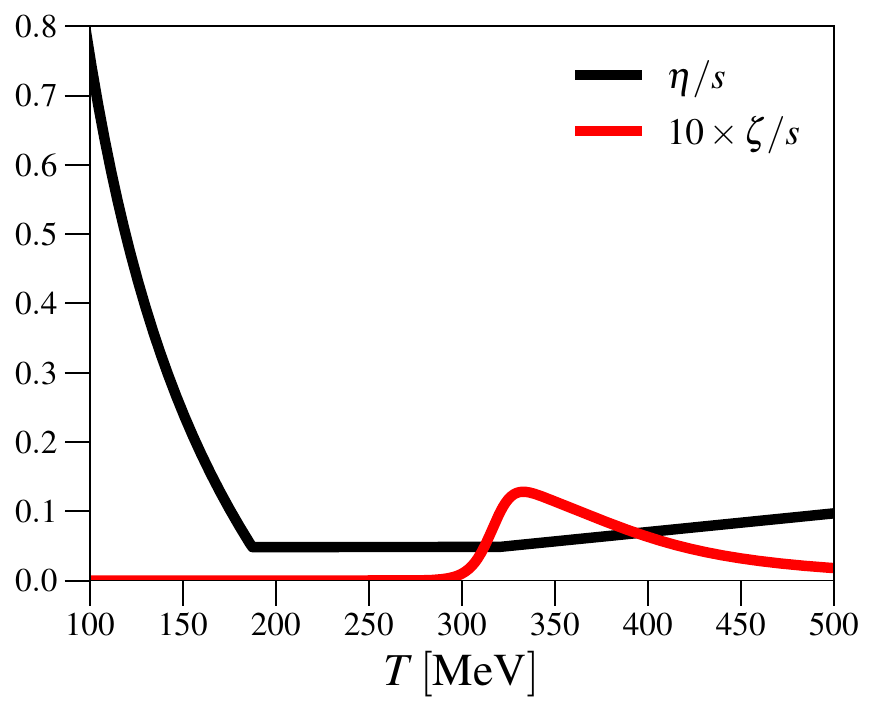}};
    \node (b) at (6.65, 0.0) {\includegraphics[width = 0.715\textwidth]{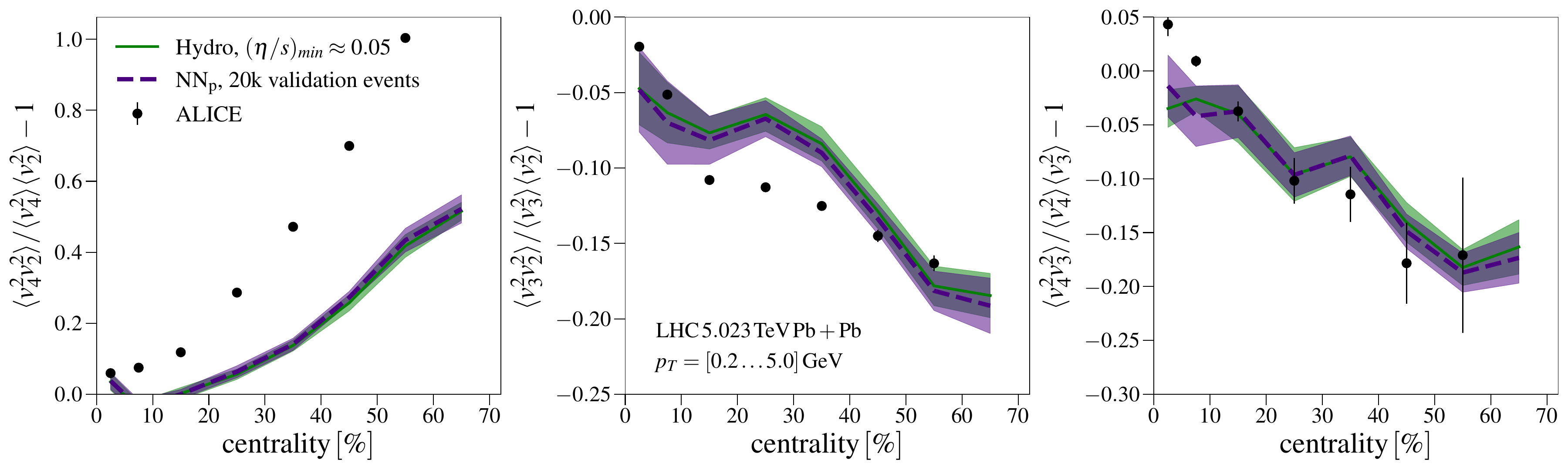}};
    \draw [ultra thick,black,->] (1.55, 0.0) to  (2.0, 0.0);
\end{tikzpicture}
\caption{Validation tests of $\rm NN_p$ networks for $NSC(m,n)$ with 20 k validation events in each case. The upper (lower) panels show result with high (low) values of specific viscosities. The experimental data are from the ALICE Collaboration \cite{ALICE:2021adw}.}
\label{fig:multi_input_validation}       
\end{figure*}

In the case of $\rm NN_p$ networks, we are mostly interested in the network accuracy for generating new events with the same parameter values as in the training data. This is because the most efficient way to do Bayesian analysis is to first generate a high number of events using neural networks in a set of parameter points, compute all of the observables in these parameter points, and then train Gaussian process emulators for these observables. Here the accuracy of $\rm NN_p$ networks was tested by taking two sets of model parameter points from the training data which correspond to drastically different values of viscosities, and then generating 20 k new independent initial state profiles for both points and doing a similar validation comparison between $\rm NN_p$ and hydrodynamic simulations as before. The results are shown in Fig.~\ref{fig:multi_input_validation}, from where one can see that $\rm NN_p$ can still very accurately reproduce the results from hydrodynamic simulations. The only exception is the peripheral region of $NSC(3,4)$ in the extremely high viscosity case, where the numerical errors of hydrodynamical simulations of themselves might be very significant.

The goal of introducing neural networks in this work was to replace the slow hydrodynamic simulations and make it possible to add multi-particle flow correlators to Bayesian analysis. We have demonstrated that this is indeed possible and has the potential to cut the computational time needed for these analyses by many orders of magnitude. 

\vspace{2mm}
\small{We acknowledge the financial support from the Jenny and Antti Wihuri Foundation, and the
Academy of Finland project 330448. This research was funded as a part of the Center of Excellence in Quark Matter of the Academy of Finland (project 346325), the European Research Council project ERC-2018-ADG-835105 YoctoLHC, and the European Union’s Horizon 2020 research and innovation program under grant agreement No 824093 (STRONG-2020). The Finnish IT Center for Science (CSC) is acknowledged for the computing time through the Project jyy2580.}

%
%
%

\end{document}